\documentclass[useAMS,usegraphicx]{mn2e}
\usepackage{rotate}
\usepackage{times}
\newif\ifAMStwofonts
\AMStwofontstrue

\def\gs{\mathrel{\hbox{\rlap{\hbox{\lower4pt\hbox{$\sim$}}}\hbox{$>$}}}}
\def\ls{\mathrel{\hbox{\rlap{\hbox{\lower4pt\hbox{$\sim$}}}\hbox{$<$}}}}

\def\Msun{M$_{\odot}$}

\def\xmm{{\it XMM-Newton}}

\def\et{{et al.\ }}

\def\pg14{{PG~1407+265}}

\def\3c{{3C~273}}

\def\rg{{\thinspace r_{\rm g}}}
\def\fvar{{F_{\rm var}}}
\def\chidof{{\chi^2_\nu/{\rm dof}}}

\def\feii{{Fe~\textsc{ii}}}

\def\A{{\rm\thinspace \AA}}
\def\cm{{\rm\thinspace cm}}
\def\erg{{\rm\thinspace erg}}
\def\eV{{\rm\thinspace eV}}

\def\ga{{\rm\thinspace gauss}}

\def\keV{{\rm\thinspace keV}}

\def\Msun{\hbox{$\rm\thinspace M_{\odot}$}}

\def\pc{{\rm\thinspace pc}}
\def\s{{\rm\thinspace s}}
\def\ks{{\rm\thinspace ks}}

\def\ph{{\rm\thinspace photons}}

\def\ergpscmps{\hbox{$\erg\cm^{-2}\s^{-1}\,$}}
\def\ergpscmpspa{\hbox{$\erg\s^{-1}\cm^{-2}\A^{-1}\,$}}
\def\ergps{\hbox{$\erg\s^{-1}\,$}}

\def\pscm{\hbox{$\cm^{-2}\,$}}

\title[X-rays from \pg14]
      {
X-rays from the radio-quiet quasar \pg14: relativistic jet or accretion
disc emission?
      }

\author[L. C. Gallo]
       {L. C. Gallo  \\
Max-Planck-Institut f\"ur extraterrestrische Physik, Postfach 1312, 85741 Garching, Germany \\
}
\date{Accepted. Received. }
\pagerange{\pageref{firstpage}--\pageref{lastpage}}
\pubyear{2005}
\begin{document}
\maketitle
\label{firstpage}

\begin{abstract}
We present two \xmm\ observations of the luminous ($L_x > 10^{46}\ergps$),
radio-quiet quasar, \pg14, separated by eleven months.
The data indicate two distinct states:
a highly variable, bright state (first epoch); and a quiescent, low-flux one
(second epoch).  During the low-flux state the spectrum is consistent with
a single, unabsorbed power law.  However, during the brighter state a 
highly variable,
steep component is statistically required.  Contemporaneous UV data from
the Optical Monitor allow an estimate of the optical-to-X-ray spectral
index ($\alpha_{ox}$), which appears typical of radio-quiet quasars during
the low-flux state, but extremely flat during the high-flux state.
The \xmm\ data can be described as originating from a combination of
jet and accretion disc processes, in which the (relativistic) X-ray jet only
works intermittently.  The scenario could help describe some of the 
complexities seen in the broadband spectral energy distribution of \pg14,
such as weak high-ionisation emission lines, strong \feii, unbeamed 
continuum, and the weak radio emission relative to the optical.

\end{abstract}



\section{Introduction}
\label{sect:intro}

The radio-quiet quasar, \pg14\ ($z=0.94$), is best known for 
constant display of enigmatic behaviour.  A detailed analysis of the
spectral energy distribution (SED; McDowell \et 1995) revealed that \pg14\
exhibited a normal non-variable, radio-quiet continuum, but weak and
blueshifted high-ionisation emission lines.  In contrast, the UV
and optical \feii\ appeared rather strong.  A hand-full of weak-line quasars
have since been discovered (e.g. Leighly \et 2004; Hall \et 2004; 
Reimers \et 2005), but no general consensus has been reached as to the
nature of the physical mechanisms involved.

A relativistically beamed continuum, as in a blazar, could explain the
absence of emission lines.  For \pg14\ this explanation is normally dismissed
as the SED of the quasar is strikingly similar to that of radio-quiet AGN
(Elvis \et 1994).  Moreover, \pg14\ is relatively radio weak with an
optical-to-radio flux ratio of $3.43$ (Kellermann \et 1989) and, by definition,
consistent with radio-quiet AGN.
However, a multi-year radio observing campaign of \pg14\ with the VLBA revealed
strong evidence of a radio jet with a highly relativistic speed
(Dopper factor $\gs 10$) and likely viewed within a few degrees of pole-on
(Blundell \et 2003).  Blundell \et suggested that the
pole-on orientation could explain some of the optical/UV properties, assuming
that the line-of-sight was into the broad line region, thus 
the emission line spectrum is diluted by the 
accretion disc continuum.

Although \pg14\ is individually curious, a better understanding of it 
could have broader implications.
The debate whether the radio characteristics
of AGN show a bimodal or continuous distribution is ongoing
(e.g. Goldschmidt \et 1999; White \et 2000; Ivezic \et 2002; 
Cirasuolo \et 2003).  The discovery of objects which possess characteristics
of both groups are obviously of interest.
It is clear that all AGN are capable of producing radio emission,
hence able to generate jets on some level.  Observational support for this
comes from sensitive, high-resolution radio imaging of radio-quiet
AGN, evidencing jets in some objects 
(e.g. Blundell \& Beasley 1998; Ulvestad \et 2005).
The simplest conclusion is that the radio mechanism in radio-quiet and
radio-loud AGN are identical, except that the bulk kinetic power is smaller
in radio-quiet objects.

To some extent, these observations form the bases of the `aborted jet' model 
which has become fashionable in AGN circles.  The idea being
that short or aborted jets are produced in radio-quiet AGN and 
the high-energy jet particles, which are expelled close to the black hole 
environment, illuminate the accretion disc  (e.g. Henri \& Petrucci 1997; 
Ghisellini \et 2004).  The presence of these energetic particles can account,
in whole or in part, for the high-energy continuum traditionally
associated with the accretion disc corona.

\pg14\ could evolve to be an important laboratory in investigation of the 
radio-quiet/loud dichotomy (or lack of it),
because the quasar clearly possesses characteristics of both classes:
it is an example of a radio-quiet AGN which has successfully launched a 
relativistic radio jet to large distances. 
In this paper, we examine the X-ray properties of this quasar to study 
the behaviour of possible jet emission closer to the central engine.

\section{Observations, data reduction, and fitting preparations}
\label{sect:data}
\pg14\ was observed with \xmm\ (Jansen et al. 2001) on 2001 January 23
(revolution 206) and 2001 December 22 (revolution 373) for approximately
71 and 42\ks, respectively.
At both epochs, all on-board instruments functioned normally.
The EPIC pn (Str\"uder et al. 2001) and MOS (MOS1 and MOS2; 
Turner \et 2001) cameras were operated in full-frame mode 
with the thin filter in place. 
The Reflection Grating Spectrometers (RGS1 and RGS2; den Herder et al. 2001)
also collected data during this time, as did the Optical Monitor 
(OM; Mason et al. 2001).  The RGS data were presented in an examination
of line-of-sight absorption toward \pg14\ (Fang \et 2005), and will not
be discussed here.

The Observation Data Files (ODFs) were processed to produce calibrated
event lists using the \xmm\ Science Analysis System ({\tt XMM-SAS v6.4.0})
and the most recent calibration files.
Due to anomalies in the Mission-Elapsed-Time during the first observation,
the default values of the the environment variable 
{\tt SAS\_OBT\_MET\_FIT}\footnote{http://xmm.vilspa.esa.es/sas/current/doc/idxEnvironments.html} were changed accordingly to ``0,1,0.5,0.9,1''.
Otherwise, standard processing practices were followed.
Unwanted hot, dead, or flickering pixels were removed as were events due to
electronic noise.  Event energies were corrected for charge-transfer
losses, and time-dependent EPIC response matrices were generated using the 
{\tt SAS} tasks
{\tt ARFGEN} and {\tt RMFGEN}.
Light curves were extracted from these event lists to search for periods of 
high background flaring.  Some flaring was evident during revolution
206, consequently the data during these intervals were simply ignored.
The source plus background photons were extracted from a
circular region with a radius of 35$^{\prime\prime}$, and the background was
selected from an off-source region with a radius of 50$^{\prime\prime}$
and appropriately scaled to the source
region.  Single and double events were selected for the pn
spectra, and single-quadruple events were selected for the MOS.
Pile-up was examined at both epochs and deemed negligible in all 
instruments.

The OM worked in imaging mode at both epochs.  In total, eight images were
taken during revolution 206, and ten during revolution 373.  Only the
$UVW2$ filter ($1800-2250\A$) was used with typical exposures between 
$3280-5000\s$.
A log of the \xmm\ observations is provided in Table~\ref{tab:log}.
\begin{table}
\begin{center}
\caption{Log of \xmm\ observations of \pg14. 
The \xmm\ revolution number of when the observation was conducted is  
given in column (1) and the observation id in column (2).  
The instrument is shown in column (3).
The total amount of useful exposure (GTI) is reported in column (4), and
the estimated number of source counts is reported in column (5).  
The combined exposure for all images is given for the OM.
The energy bands considered are $0.25-10\keV$ and $0.3-10\keV$ 
for the pn and MOS, respectively.
}
\begin{tabular}{ccccc}                
\hline
(1) & (2) & (3) & (4) & (5) \\
Rev. & Obs. ID & Instrument & Exposure (\s) & Counts \\
\hline
             206 & 0092850101 &  pn    & 51458 & 125750 \\
                &  &  MOS1  & 62925 & 35331 \\
                &  &  MOS2  & 62821 & 35419 \\
                &  &  OM    & 36000 & --    \\
\hline
             373 & 0092850501 &  pn    & 35098 & 30980 \\
                 &  &  MOS1  & 41017 & 8979  \\
                 & &  MOS2  & 41007 & 8784  \\
                 & &  OM    & 38800 & --    \\
\hline
\label{tab:log}
\end{tabular}
\end{center}
\end{table}

In the following analysis, the spectra are grouped such that each bin 
contains at least 20 counts, and fitting is performed using 
{\tt XSPEC v12.2.0} (Arnaud 1996).
Fit parameters are reported in the rest frame of the object and
quoted errors correspond to a 90\% confidence
level for one interesting parameter (i.e. $\Delta\chi^2$ = 2.7 criterion).
K-corrected luminosities are derived assuming isotropic emission,
a value for the Hubble constant of $H_0$=$\rm 70\ km\ s^{-1}\ Mpc^{-1}$, and
a standard cosmology with $\Omega_{M}$ = 0.3 and $\Omega_\Lambda$ = 0.7.

A value for the Galactic column density toward \pg14\ of 
$1.38 \times 10^{20}\pscm$ (Elvis \et 1989) is adopted in all of the 
spectral fits.
Initial spectral fits were performed with a free absorption parameter set to 
the redshift of the source.  In determining that the intrinsic absorption
was imperceptible ($\ls 5\times10^{19}\pscm$), 
subsequent spectral modelling included only the
Galactic component.

Recent work on the pn calibration show that the spectral response is reasonable
down to $0.25\keV$ (F. Haberl priv. comm.).  We confirmed that inclusion of
the $0.25-0.3\keV$ data did not hamper the fit; therefore we elected to 
utilise the lower energy range in our analysis.

On the low-energy end, the MOS are limited to $0.3\keV$ to minimise 
cross-calibration uncertainties between the three cameras (Kirsch 2005).
At high energies, all three spectra are restricted at $10\keV$ where
background contamination becomes significant.

\section{The bright-state January 2001 observation}
\label{sect:jan}

\subsection{Spectral analysis}
\label{sect:jansa}

Fitting the pn and MOS spectra separately revealed relatively good agreement
within known cross-calibration uncertainties (Kirsch 2005); therefore
for the initial phenomenological fits
all of the EPIC data were fitted together, while
residuals from each instrument were examined separately to distinguish
gross discrepancies.
A moderately steep power law ($\Gamma = 2.32\pm0.02$) fitted the observed 
$1.3-10\keV$ band ($2.5-19.4\keV$ rest frame) relatively well
($\chidof = 0.96/699$).   A search for a narrow 
($\sigma < 150\eV$) iron emission line between $6-7\keV$ yielded no detection,
with the maximum equivalent width of any such feature being $EW < 12\eV$.

\begin{figure}
\rotatebox{270}
{\scalebox{0.32}{\includegraphics{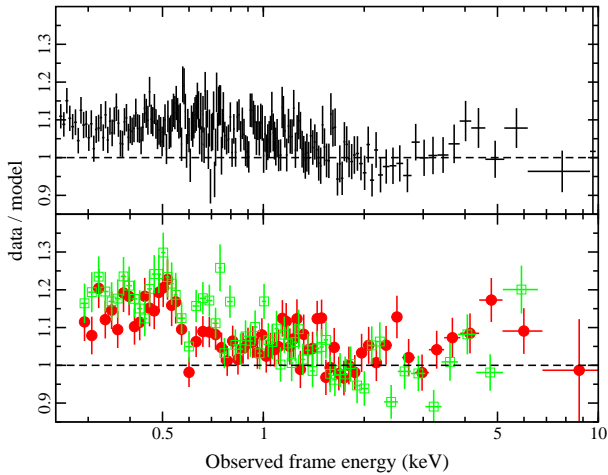}}}
\caption{The residuals remaining in the pn (top panel) and MOS (lower panel;
MOS1 red, filled circles; MOS2 green, open squares)
data from revolution 206,
when extrapolating the $1.3-10\keV$ band power law ($\Gamma \approx 2.32$)
to lower energies.  The data are binned for display purposes.
}
\label{fig:resid}
\end{figure}
Extrapolation of the power law fit to $0.25\keV$ in the pn and $0.3\keV$ in
the MOS degraded the fit quality ($\chidof = 2.84/1047$),
and showed a gradual excess in the residuals toward lower energies
(Figure~\ref{fig:resid}).
Several models were attempted to
describe the broadband ($\sim 0.49-19.4\keV$ rest frame) spectra of \pg14\ in a
phenomenological manner (Table~\ref{tab:fit206}).
\begin{table}
\begin{center}
\caption{Combined fit to all of the EPIC data of \pg14\ during
revolution 206.
The model, quality of fit, and rest frame model parameters are given in
columns (1), (2), and (3), respectively.
}
\begin{tabular}{ccc}                
\hline
(1) & (2) & (3) \\
Model & $\chidof$ & Parameters \\
\hline
power law      & $1.11/1047$  & $\Gamma=2.40\pm0.01$ \\
\hline
broken         & $1.07/1045$  & $\Gamma_1 = 2.41\pm0.01$  \\
power law      &              & $E_b = 4.43^{+0.92}_{-0.80}\keV$ \\
               &              & $\Gamma_2 = 2.27^{+0.04}_{-0.06}$ \\
\hline
blackbody      & $1.09/1045$  & $kT = 204^{+28}_{-32}\eV$ \\
plus           &              & $\Gamma = 2.36\pm0.01$ \\
power law      &              &  \\
\hline
double         & $1.08/1045$  & $\Gamma_1 = 2.47\pm0.02$ \\
power law      &              & $\Gamma_2 = 1.75\pm0.18$ \\
\hline
\label{tab:fit206}
\end{tabular}
\end{center}
\end{table}

First, the single power law was formally fitted to the broadband spectra.
This was an improvement over the extrapolated, high-energy power law,
but, predictably, the data at higher energies (where statistics are 
lower) were underestimated.

Broken power law, double power law, and blackbody plus power law models,
all provided comparable improvement over the single power law fit, and 
the inclusion of a second continuum component (two additional free parameters)
was a significant improvement.  For example, when tested with an F-test,
the broken power law model resulted in an F-test probability of $10^{-9}$.
We note that the double power law fit presented in Table~\ref{tab:fit206}
was not unique.  Different initial parameters resulted in slightly
different final parameters, but comparable $\chi^2_{\nu}$.

Remarkable was the high energy where the apparent change in spectral 
slope occurs in all of the models.  This is seen in the high break energy
($E_b \approx 4.43\keV$ rest frame) in the broken power law, and also
manifested in the high temperature of the blackbody component 
($kT \approx 204\eV$).  As in the broken power law fit, 
the steeper component dominates the spectrum up to 
rather high energies in the double power law model, as well.
It is unlikely that the high temperature or high break energy are
physically meaningful, rather they instead indicate the presence of another
component, which cannot be uniquely determine with the current statistics
(see Sect.~\ref{sect:phnat}).

Considering the broken power law fit, the observed $0.3-10\keV$ flux, corrected
for Galactic extinction, was about $5\times 10^{-12}\ergpscmps$.  The 
corresponding rest frame luminosities in the $0.3-2$ and $2-10\keV$ range
were $2.14\times 10^{46}$, and $8.85\times 10^{45}\ergps$, respectively.
The $2-10\keV$ luminosity was comparable to the 1993 measurement made from
an $ASCA$ observation ($L_{2-10} \approx 7.5\times10^{45}$, 
Reeves \& Turner 2000).

\subsection{Variability }

\subsubsection{Broadband X-ray flux variability }

\begin{figure}
\rotatebox{270}
{\scalebox{0.32}{\includegraphics{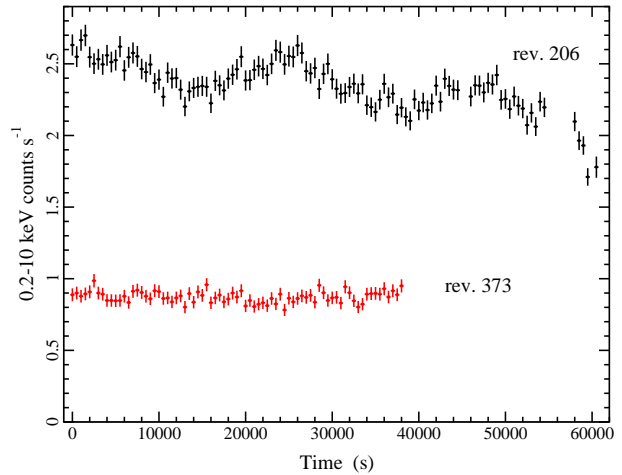}}}
\caption{The $0.2-10\keV$ pn light curves of \pg14\ during the January 2001
(rev. 206) observation (top curve) and December 2001 (rev. 373) observation
(bottom curve).  There is clear variability during the brighter observations,
whereas the light curve is consistent with a constant during revolution 373.
The data are binned in $500\s$, and $0\s$ on the time axis marks the start of 
each observation.
}
\label{fig:lc}
\end{figure}

Examining the observed $0.2-10\keV$ light curve in $500\s$ bins 
revealed significant variations on the order of $\pm20\%$ about the mean count
rate over the duration of the revolution 206 observation (see 
Figure~\ref{fig:lc}).  The variability is even more impressive when we
recall that the average luminosity of \pg14\ was in excess of $10^{46}\ergps$, 
and that the quasar is radio-quiet.

The most rapid rise occurred about $17\ks$ into the observations when the
count rate increased about $10\%$ in $\sim 9\ks$.  
Adopting the blackbody plus power law fit resulted in the most conservative
estimate of the luminosity rate of change.  Accordingly, the change in count rate
corresponds to a luminosity change of 
$\Delta L \approx 2.84 \times 10^{45}\ergps$ in a quasar rest frame time
interval of $\Delta t \approx 4600\s$.
Following Fabian (1979; see also Brandt \et 1999) the radiative
efficiency of \pg14\ was $\eta \gs 0.29$.  Assuming a uniform, spherical
emission region the measured $\eta$ is comparable to
the efficiency expected from a maximally rotating Kerr black hole
($\eta \approx 0.3$, Thorne 1974).  The demand on black hole spin
can be relaxed if asymmetric emission processes, such as relativistic Doppler 
boosting or light bending, play an important role in the energy output
of \pg14.

\subsubsection{Interband variability }
\label{sect:206iv}

Initially, variability was examined in three sub-bands: $0.2-0.7$,
$0.7-2$, and $2-10\keV$.  It was established that there was no discernible
spectral variability or significant differences in the amplitude of the 
fluctuations between the $0.7-2\keV$ and $2-10\keV$ bands, thus to
improve statistics, the two bands were combined.  Therefore, 
light curves are examined in the $0.2-0.7\keV$ ($0.4-1.35\keV$ rest frame) 
and $0.7-10\keV$ ($1.35-19.4\keV$ rest frame) bands.

It was immediately noticeable that the degree of variability was much greater
in the lower energy band.  The fractional variability ($\fvar$) in the
$0.2-0.7\keV$ band was $7.33\pm0.64\%$, whereas between $0.7-10\keV$,
$\fvar = 3.34\pm0.73\%$ (uncertainties estimated following Edelson \et 2002).

\begin{figure}
\rotatebox{270}
{\scalebox{0.32}{\includegraphics{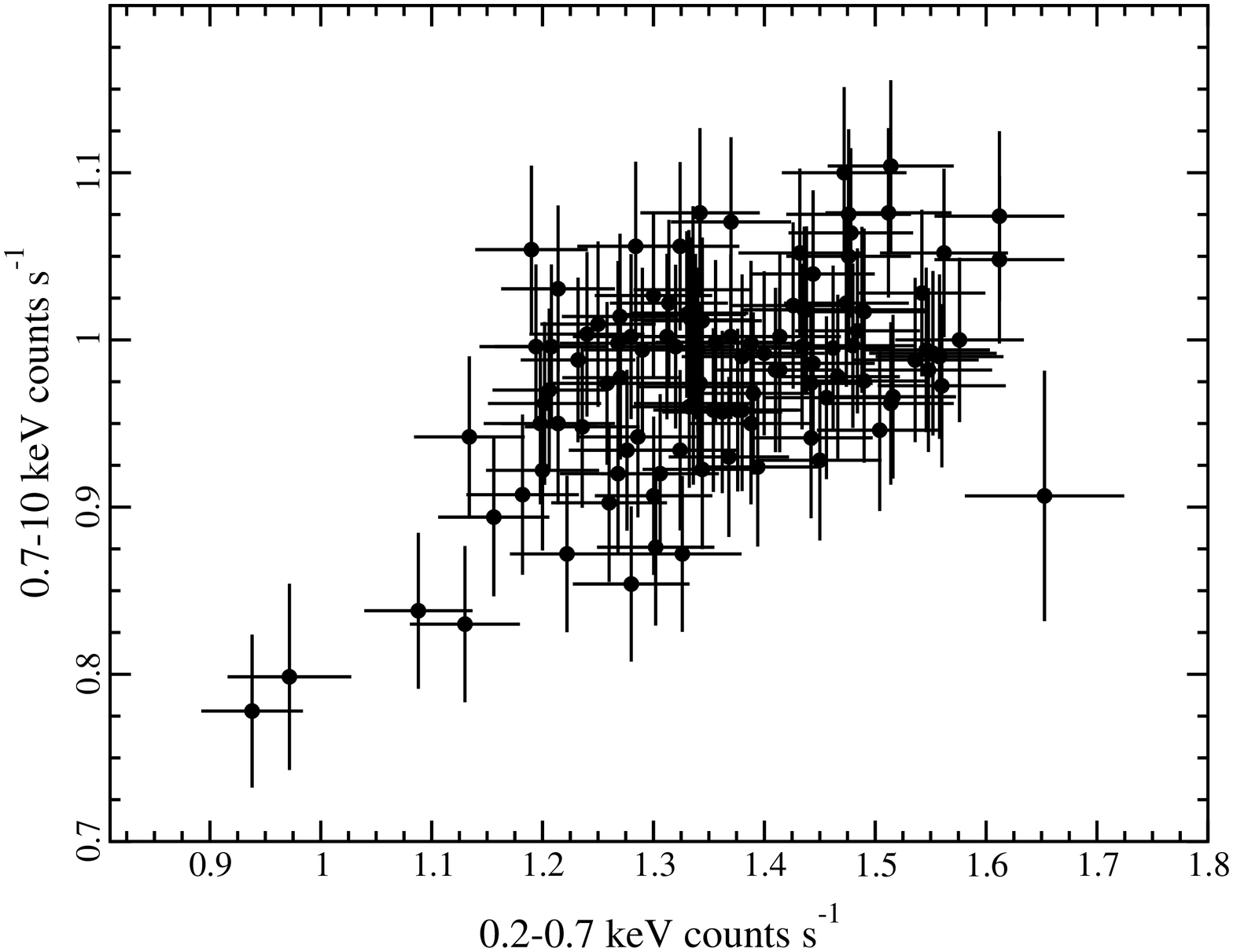}}
\scalebox{0.32}{\includegraphics{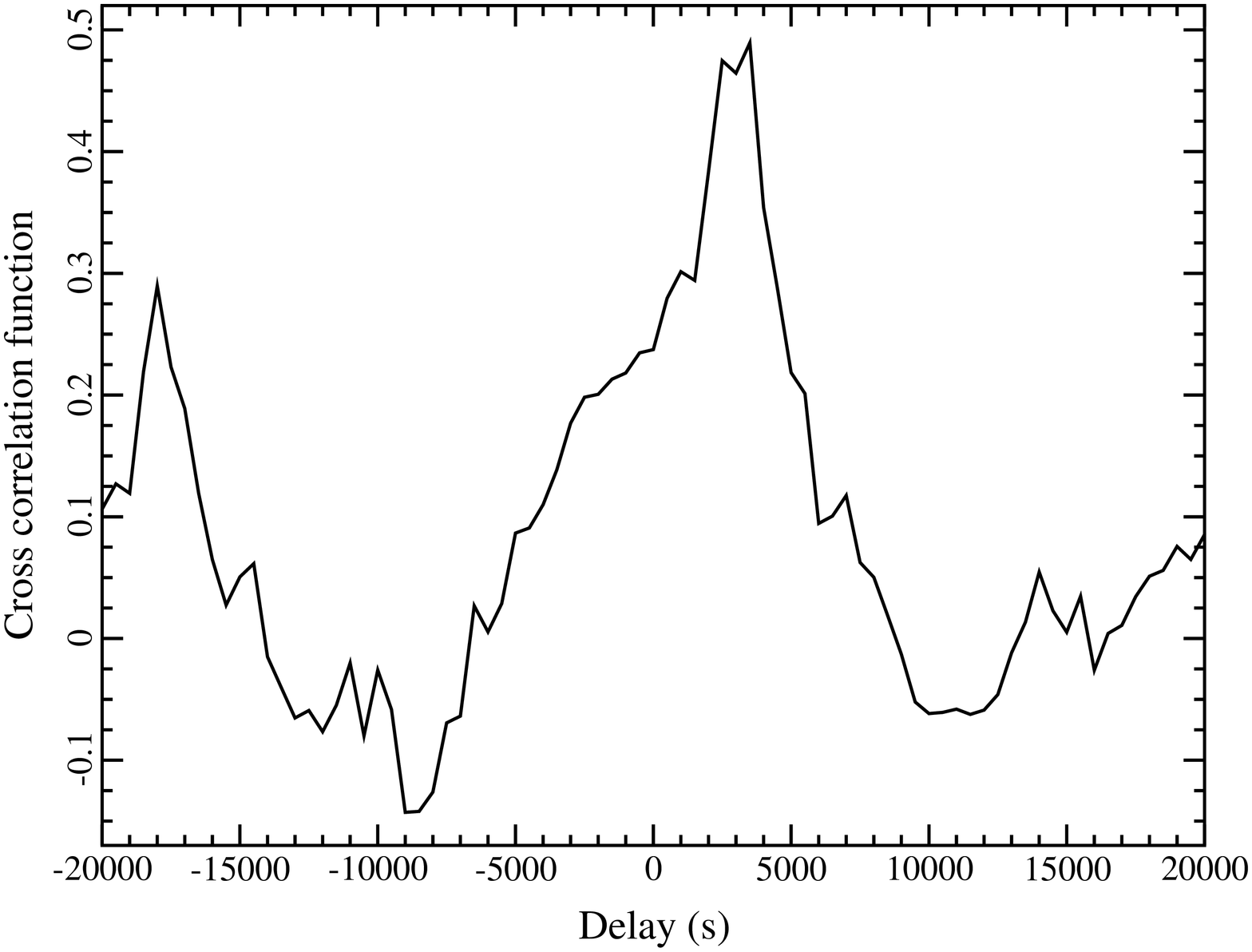}}}
\caption{The count rate correlation between the $0.7-10\keV$ and $0.2-0.7\keV$
bands is shown in the top panel.  The cross-correlation function 
between the two bands is shown in the bottom panel.  The positive offset 
indicates that the hard band follows the soft band by about $3000\s$.
The data are binned in $500\s$.
}
\label{fig:ccf}
\end{figure}
The count rates measured in the two bands were well-correlated 
(Figure~\ref{fig:ccf}).  As such, we examined possible lags between
the two bands by calculating the nominal cross-correlation function.
With the light curves in $500\s$ bins, there does appear to be a
lag (Figure~\ref{fig:ccf}), such that the low-energy band leads the
hard energy band by $\sim 3000\s$.

The significance of the lag should be treated with caution.  When
we cross-correlated the $0.2-0.7\keV$ band with $0.7-2\keV$ and $2-10\keV$
light curves separately, no lag was detected.  It could simply be that the
combined $0.7-10\keV$ light curve produces sufficiently high signal-to-noise
to make a detection, but nevertheless we do not consider the possible lag a 
robust detection.

\subsubsection{X-ray spectral variability }
\label{sect:206hilo}

\begin{figure}
\rotatebox{270}
{\scalebox{0.32}{\includegraphics{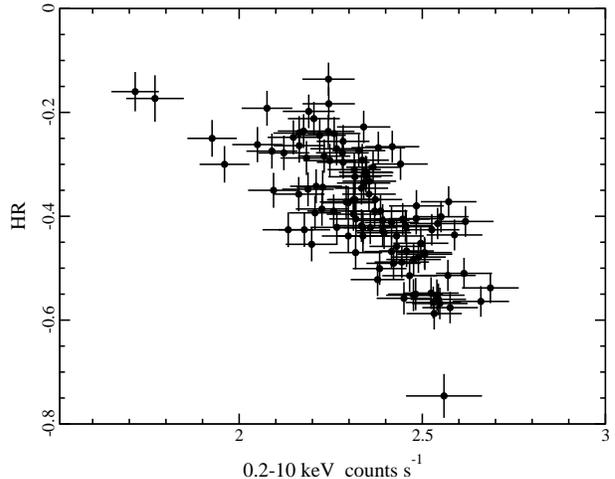}}}
\caption{The hardness ratio ($HR$) as a function of $0.2-10\keV$ count rate
during revolution 206.  The spectrum becomes softer as the intensity 
increases.  The pn data are binned in $500\s$ intervals.
}
\label{fig:hrcr}
\end{figure}
As discussed in Section~\ref{sect:206iv} there was clear spectral
variability with the amplitude of the fluctuations being much larger
in the $0.2-0.7\keV$ band then in the $0.7-10\keV$ band.  We investigated
the spectral variability with time by calculating the hardness ratio
($HR = H-S/H+S$, where $H$ and $S$ are count rates in the hard and soft
bands, respectively) variability curve.  The hardness ratio curve was
inconsistent with a constant ($\chi^2_\nu = 1.58$) and correlated with
the broadband count rate, such that the spectrum becomes harder with 
diminishing intensity (Figure~\ref{fig:hrcr}).  In general, as the flux
gradually decreased over the $\sim70\ks$ observation, the spectrum became
harder.
To investigate this short-term, flux-dependent spectral variability during
revolution 206, we divided the observation into half: the high-flux
state (first half) and a low-flux state (second half).  

\begin{figure}
\rotatebox{270}
{\scalebox{0.32}{\includegraphics{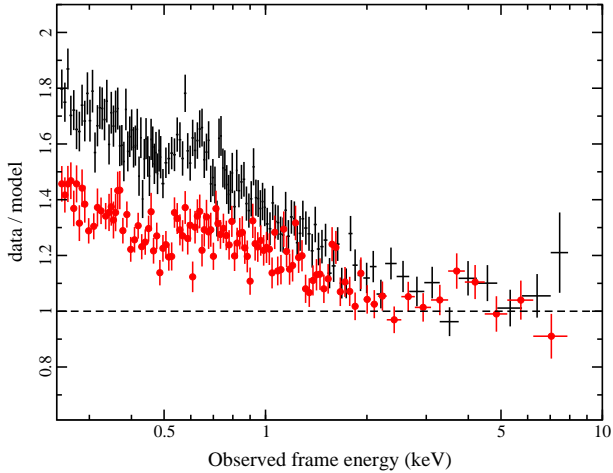}}}
\caption{The ratio from extrapolating a $3-10\keV$ power-law 
($\Gamma\approx 2.25$) to lower energies for the high-flux (black crosses)
and low-flux (red dots) pn spectra during revolution 206.  The data are
binned for display purposes.
}
\label{fig:206hilo}
\end{figure}

Initially, we fitted both pn spectra with a power law above $3\keV$ only.
We found that this produced a good fit with a common flux and photon
index ($\Gamma\approx 2.25$), indicating that the spectral variability
was dominated by the lower-energy component.  
Extrapolating this power law to lower
energies, we determine that the high-flux spectrum appeared significantly
steeper (Figure~\ref{fig:206hilo}).

To investigate this further, we fitted both spectra simultaneously with
a broken power law.  We maintained the break energy ($E_b$) and
high-energy power law ($\Gamma_2$) common for both models and allowed only
the low-energy power law to vary separately.
This produced a good fit to the data ($\chi^2_{\nu} = 1.02$) and illustrated
the nature of the spectral variability (see fit parameters in 
Table~\ref{tab:206hilo}): the high-energy spectrum was consistent in
flux and spectral shape; however, at low energies the variability
seemed to be characterised by a change in shape as well as flux.
\begin{table}
\begin{center}
\caption{A broken power-law fit to the high- and low-flux spectra
of \pg14\ during revolution 206.
The energy band utilised is $0.25-10\keV$
($0.49-19.4\keV$ in the rest frame).
The model parameter is shown in column (1), $n$ is the model normalisation at $1\keV$
in units of $10^{-4} \ph \keV^{-1} \cm^{-2} \s^{-1}$.  
The values during the high-flux
and low-flux state are given in columns (2) and (3), respectively.
$E_b$ and $\Gamma_2$ are fixed between the high and low state.
}
\begin{tabular}{ccc}                
\hline
(1) & (2) & (3) \\
Model & High-flux & Low-flux \\
Parameter & & \\
\hline
$\Gamma_1$     & $2.49\pm0.01$           & $2.39\pm0.01$ \\
$E_b$ (keV)   & $6.14^{+1.54}_{-1.52}$  & $6.14^{+1.54}_{-1.52}$  \\
$\Gamma_2$     & $2.27^{+0.09}_{-0.11}$  & $2.27^{+0.09}_{-0.11}$ \\
$n$            & $10.4\pm 0.1$           & $9.7\pm 0.1$ \\
\hline
\hline
\label{tab:206hilo}
\end{tabular}
\end{center}
\end{table}

\subsubsection{Possible ultraviolet variability }

Eight images of \pg14\ were obtained in the $UVW2$ filter (rest frame
$930-1160\A$ range) with the OM during the first $35\ks$ of the X-ray
observation.  For each image, a flux density was estimated from the count rate
using the conversion factor derived by Chen (2004).
The average $UVW2$ flux density during revolution 206 was approximately
$7.54\times10^{-15}\ergpscmpspa$.

\begin{figure}
\rotatebox{270}
{\scalebox{0.32}{\includegraphics{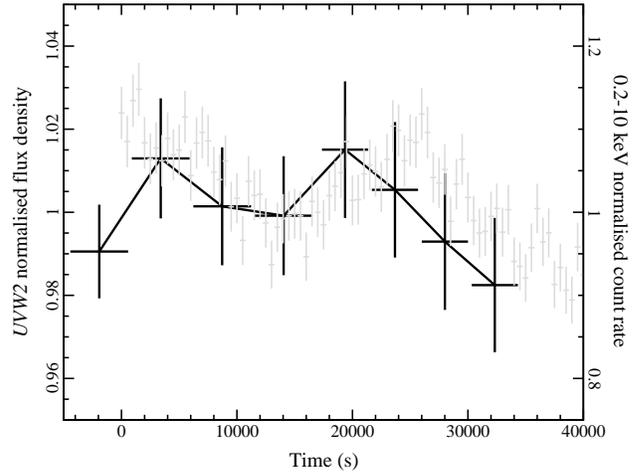}}}
\caption{The $UVW2$ normalised light curve overplotted (in time) on a segment of the 
$0.2-10\keV$ normalised light curve during revolution 206.
The left axis correspond to the UV data and the right axis to the X-rays.
The $0\s$ on the time axis marks the start of the X-ray observation.
The possible common trend could be strengthened by a slight time shift in the
light curves.
N.B. The figure should be considered illustrative, as
the fluctuations in the UV are really much less then in the X-rays (note the 
values on the respective axes).
}
\label{fig:xuv}
\end{figure}

We emphasise that within the flux uncertainties ($\sim 5\%$) there were
no significant variations in the UV light curve.  
Formally, {\em the OM light curve was consistent
with a constant}.  However, the apparent trend in the light curve was 
intriguing.  Taking liberty to speculate, when the $UV$ light curve was 
overplotted (in time) on the X-ray light curve there appeared to be a common 
trend
in the fluctuation (Figure~\ref{fig:xuv}).  The relationship between the
curves could even be improved if we allowed for a slight time shift in
one of them (e.g. if the UV lead the X-rays by a few kiloseconds).
Admittedly, Figure~\ref{fig:xuv} can 
appear deceiving, and we reiterate that the UV fluctuations are on the level
of $\ls5\%$, whereas the X-rays variations are about $30\%$.  Therefore,
Figure~\ref{fig:xuv} should be considered illustrative, in absence of higher
quality UV data.

\section{The low-flux December 2001 observation}
\label{sect:dec}

\subsection{Spectral analysis}

\begin{table}
\begin{center}
\caption{Combined fit to all of the EPIC data of \pg14\ during
revolution 373.  The energy band utilised is $0.25-10\keV$
($0.49-19.4\keV$ in the rest frame).
The model, quality of fit, and rest frame model parameters are given in
columns (1), (2), and (3), respectively.
}
\begin{tabular}{ccc}                
\hline
(1) & (2) & (3) \\
Model & $\chidof$ & Parameters \\
\hline
power law      & $0.99/830$  & $\Gamma=2.21\pm0.01$ \\
\hline
broken         & $0.98/828$  & $\Gamma_1 = 2.36^{+0.10}_{-0.04}$ \\
power law      &             & $E_b = 1.30\pm0.19\keV$ \\
               &             & $\Gamma_2 = 2.19\pm0.02$ \\
\hline
blackbody      & $0.97/828$  & $kT = 108\pm13\eV$  \\
plus           &             & $\Gamma = 2.19\pm0.02$ \\
power law      &             &  \\
\hline
double         & $0.98/828$  & $\Gamma_1 = 2.67^{+0.33}_{-0.90}$ \\
power law      &             & $\Gamma_2 = 2.11^{+0.08}_{-0.15}$ \\
\hline
\label{tab:fit373}
\end{tabular}
\end{center}
\end{table}
A power law ($\Gamma = 2.19\pm0.04$) provided an acceptable fit 
($\chidof = 0.90/454$ to the
$1.3-10\keV$ ($2.5-19.4\keV$ rest frame) EPIC spectra of \pg14
during revolution 373.  As with the earlier observation, no 
narrow iron emission line was detected.

Extrapolating the high-energy power law to lower energies was a good
approximation to the broadband spectra.  Indeed, fitting the single power
law over this entire band did not result in a significant change of the
photon index ($\Delta\Gamma \approx 0.02$).
The addition of a second continuum component did improve the fit
over the single power law (F-test probability of $\approx 10^{-4}$), but
not as significantly as in the revolution 206 data.

Adopting the broken power law fit, resulted in an unabsorbed
$0.3-10\keV$ flux of about $2\times 10^{-12}\ergpscmps$.  The
corresponding rest frame luminosities in the $0.3-2$ and $2-10\keV$ range
were $7.29\times 10^{45}$, and $3.95\times 10^{45}\ergps$, respectively.

\subsection{Timing analysis}
The timing behaviour of \pg14\ during revolution 373 was much more in-line
with the expected behaviour of a high-luminosity quasar.
The $0.2-10\keV$ light curve (Figure~\ref{fig:lc}) was completely consistent
with a constant over the $\sim40\ks$ observations ($\chidof=0.81/76$).
Similarly, the UV light curve was also constant.

The X-ray spectral variability was also negligible.  The fractional
variability in the two X-ray sub-bands (i.e. $0.2-0.7\keV$ and $0.7-10\keV$)
was on the $3\%$ level, and the hardness ratio was
also rather constant over time ($\chidof=1.11/76$).

\section{Comparison between the two observations}
\label{sect:comp}

\subsection{Optical-to-X-ray spectral index}

The $UVW2$ flux density of \pg14\ diminished 
from $(7.54\pm0.03) \times 10^{-15}\ergpscmpspa$ during revolution 206
to $(6.82\pm0.05) \times 10^{-15}\ergpscmpspa$ during revolution 373.
The fractional decrease in the AGN UV flux density was about $10\%$
over the eleven month period.  
Over the same time, the average broadband X-ray flux diminished by 
$\sim60\%$.  

A serendipitous, optically bright object ($UVW2\approx 13.51$) was 
identified in the OM
field-of-view in two images at each epoch.  The object is possibly 
associated with the F8 star HD~124732, but this is not conclusive given
the several arc minute discrepancy in positions.  Of importance was that
the flux of the object ($2.01\times 10^{-14}\ergpscmpspa$) was equal at both 
epochs (revolution 206 and 373) to within about $1\%$. 
This clearly indicates that the different UV fluxes measured in \pg14\
arise from changes in the AGN and not from instrumental
effects of the OM. 

The optical-to-X-ray spectral index, $\alpha_{ox}$, is the slope of a 
hypothetical power law extending between the optical (UV) and X-ray
continuum in AGN:
$
\alpha_{ox} = log(f_{x}/f_{o})/log(\nu_{x}/\nu_{o}).
$
In the standard definition $f_{x}$ and $f_{o}$ are the intrinsic flux 
densities at $2\keV$ and $2500\A$, respectively, consequently
$
\alpha_{ox}^{\prime} = 0.384 log(f_{2keV}/f_{2500\A}).
$

Typical values of $\alpha_{ox}^{\prime}$ for unabsorbed, radio-quiet AGN
are between $-1.2$ and $-1.8$ with the slope steepening with increasing
UV luminosity (e.g. Strateva \et 2005).  In radio-quiet AGN, 
$\alpha_{ox}^{\prime}$ is likely representative of accretion disc processes,
and understanding the origin and distribution of $\alpha_{ox}^{\prime}$
characteristics will lead to a bettering understanding of the primary emission 
mechanism in AGN.

With the data at hand, we were able to measure the spectral index of the
power law between $2\keV$ and $1090\A$ (the rest frame peak emission observed
in the $UVW2$ filter):  
$
\alpha_{ox} = 0.445 log(f_{2keV}/f_{1090\A}).
$
The spectral slope between $1090\A$ and $2500\A$ is then:
$
\alpha_u = 2.783 log(f_{1090\A}/f_{2500\A}).
$
From these equations we can determine a transformation from our measured
$\alpha_{ox}$ and the standard definition, $\alpha_{ox}^{\prime}$,
specifically:
$
\alpha_{ox}^{\prime} = 0.138\alpha_u + 0.863\alpha_{ox}.
$ 

This leaves us to estimate $\alpha_u$ and thus requires that we make some 
assumptions.  Firstly, we assume that the UV $continuum$ properties of 
\pg14\ are not unusual.  While it is clear that the emission line properties
are unique, McDowell \et (1995) concluded that the continuum was rather typical
of radio-quiet AGN.  As such, we adopt the spectral slope of the composite 
SDSS quasar spectrum between $1300-5000\A$ ($\alpha_u = -0.44$,
Vanden Berk \et 2001).  This brings us to the second assumption, that
the slope between $1090\A$ and $1300\A$ is not drastically different.
This is likely invalid (e.g. Laor \et 1995; Shang \et 2005), but since the 
systematic uncertainty will be introduced to both measurements, the differences
between the two epochs should remain precise.

During the low-flux observation (revolution 373), we measured 
$\alpha_{ox}^{\prime} \approx -1.23$, which is not atypical of unabsorbed, 
radio-quiet
AGN (e.g. Strateva \et 2005; Yuan \et 1998; Elvis \et 1994).  Interestingly,
during the high-flux, revolution 206 observation, $\alpha_{ox}^{\prime} \approx -1.09$,
very flat, even in terms of BL~Lac spectral indices (e.g. Perlman \et 2005).

We recalculate $\alpha_{ox}^{\prime}$ during revolution 206, but this time we 
considered only the $2\keV$ emission from the hard power law component of the
spectrum.  This time, $\alpha_{ox}^{\prime} \approx -1.39$, much more comparable to the
revolution 373 measurement.  This may be implying that the high energy 
component originates from a less variable, accretion dominated process, whereas
the steep, soft component may be beamed emission.

\subsection{Long-term X-ray variability}

\begin{figure}
\rotatebox{270}
{\scalebox{0.32}{\includegraphics{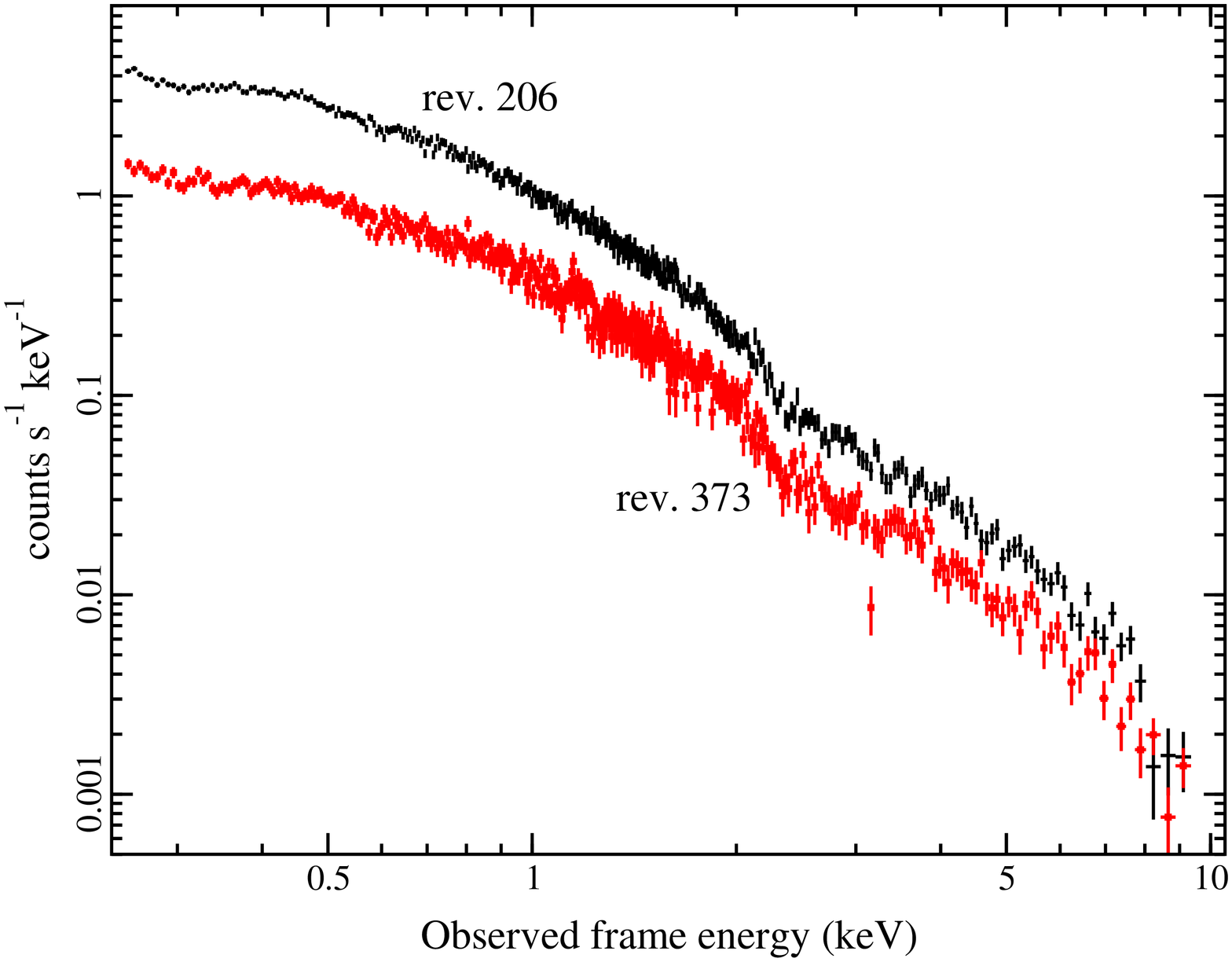}}
\scalebox{0.32}{\includegraphics{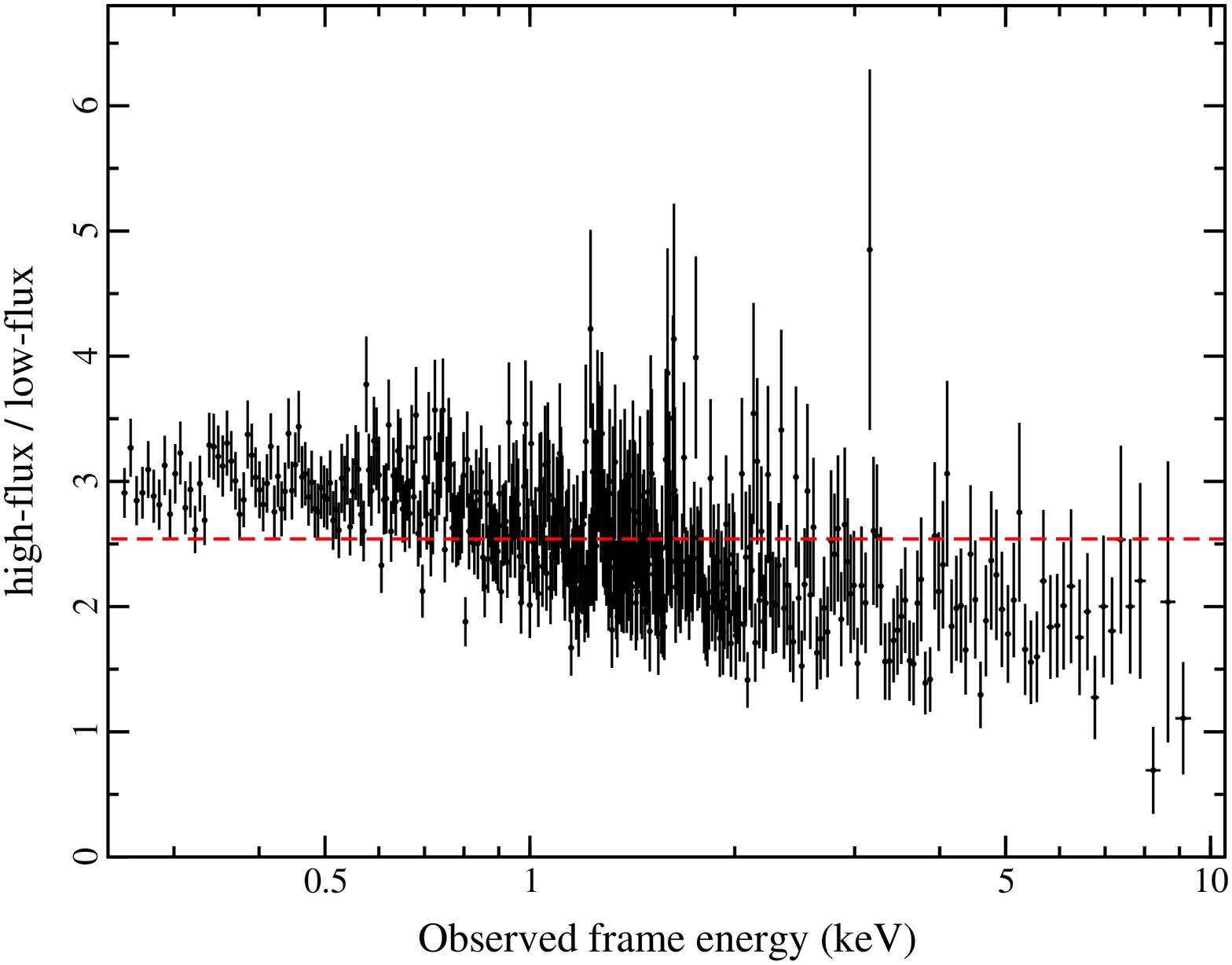}}
\scalebox{0.32}{\includegraphics{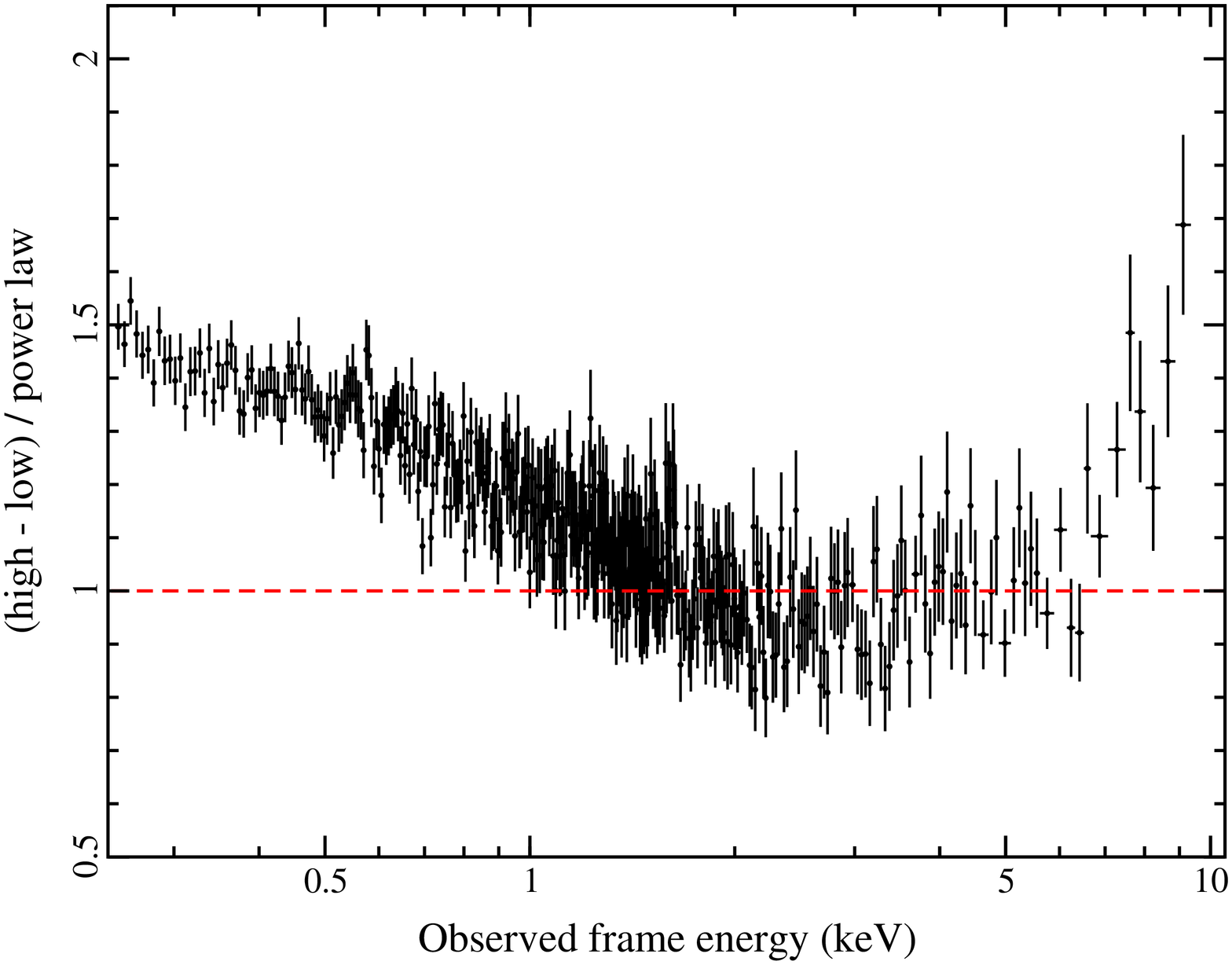}}}
\caption{In the top panel the $0.25-10\keV$ pn count spectra of \pg14\ during 
January 2001 (rev. 206; top curve) and December 2001 (rev. 373; bottom curve)
are displayed.  In the middle panel is the ratio spectrum of the high-flux
(rev. 206) data over the low-flux (rev. 373) data.  The dashed line marks the
average flux change between $0.25-10\keV$.  The flux increases more 
dramatically as energy decreases.
The bottom panel displays the ratio of the difference spectrum (high - low)
fitted with a power law ($\Gamma = 2.18\pm0.03$) in the rest frame range
$2.5-19.4\keV$ ($1.3-10\keV$ in the observed frame).  The difference spectrum
has been corrected for Galactic extinction, and clearly deviates from a 
power law at high and low energies.
}
\label{fig:xdiffs}
\end{figure}

Considering its high luminosity and radio-quiet nature, the X-ray spectrum of
\pg14\ displayed impressive spectral changes over an eleven month period.
In Figure~\ref{fig:xdiffs} some of the X-ray spectral difference between
the two observational epochs are portrayed.  A ratio spectrum between the
high-flux (rev. 206) and low-flux (rev. 373) data illustrate how the 
flux difference between the two epochs is not uniform across the spectrum,
but gradually increases toward lower energies.

The difference spectrum (high $-$ low) also indicates complex spectral changes.
A power law ($\Gamma = 2.18\pm0.03$), corrected for Galactic absorption,
and fitted to the difference spectrum between the rest frame $2.5-19.4\keV$ 
($1.3-10\keV$ in the observed frame) band gives a good result.
This high-energy band was selected as it appears to show the least amount
of spectral variability according to the analysis in Sect.~\ref{sect:jan} and
\ref{sect:dec}.  However, there are some deviations at the highest energies.
On extrapolating the power law to lower energies significant
deviations are seen (Figure~\ref{fig:xdiffs}), indicating that there are
changes in the spectral shape.  This is supported by the different 
phenomenological fits found when fitting the two data sets separately
(see Table~\ref{tab:fit206} and \ref{tab:fit373}).

The residuals in the difference spectrum
imply changes in the shape of (at least) one spectral component. 
The fact that the residuals form an excess at higher and lower energies
may suggest that if the variations are due to one spectral component, this
component is likely not smooth in nature (like a power law), but 
may have some curvature (as may be expected from jet emission).

\subsection{Flux-flux plot analysis}

\begin{figure}
\rotatebox{270}
{\scalebox{0.32}{\includegraphics{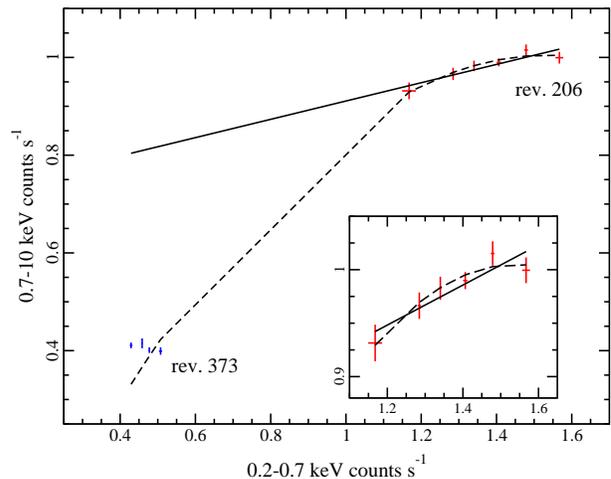}}}
\caption{The flux-flux plot for the $0.2-0.7\keV$ and $0.7-10\keV$
light curves.  The higher flux (revolution 206) data can be well
approximated by a linear (solid line) or a quadratic (dashed curve) 
fit (shown in detail in the
inset).  On extrapolating the quadratic and linear fits to lower
count rates, we find that the quadratic model predicts the lower-flux
(revolution 373) behaviour better.
}
\label{fig:ff}
\end{figure}

We further investigated the spectral variability by considering the
correlation in the flux variations in the two energy bands used.
Taylor \et (2003) demonstrated that the flux-flux plot can be effectively
used for this purpose.  If the varying component of a spectrum has a constant
shape (i.e. variations are due to changes in normalisation only) then
the flux-flux relation will be linear.  On the other hand, the 
relation will not be linear if the spectral shape also changes.

We constructed a flux-flux plot (Figure~\ref{fig:ff}) as demonstrated by 
Taylor \et (2003).
In considering the variable high-flux observation, both a linear and quadratic
function fit the data equally well ($\chi^2_{\nu} = 0.93$ and $0.37$, 
respectively), although the quadratic fit does seem a better approximation 
to the data (see the inset in Figure~\ref{fig:ff}).

The possible non-linearity displayed in the flux-flux plot during revolution
206 implies that the short-term (within the observation) variations are not
due to a simple change in normalisation of one component, but require
a change in the spectral shape as well (i.e. pivoting of the power law at
high energies; e.g. Zdziarski \et 2002).  This is 
consistent with the flux-dependent spectral variability seen in
Figure~\ref{fig:hrcr} and the spectral differences seen in the analyses of
the high- and low-flux state spectra during revolution 206 (see 
Figure~\ref{fig:206hilo}, Table~\ref{tab:206hilo}, and 
Section~\ref{sect:206hilo}).

Extrapolation of the linear and quadratic fits to lower count rates in the
flux-flux plot showed that the quadratic fit also predicted the low-flux
state of revolution 373 much better than the linear fit (Figure~\ref{fig:ff}).
As with the difference spectrum (Figure~\ref{fig:xdiffs}), this
further emphasises that the long-term variability, which is only examined at
two epochs separated by eleven months, required changes in the shape of
the variable component.

\section{Discussion}

\subsection{General Findings}

The main results of this analysis are the following:

\begin{itemize}
\item[(1)]
In comparing X-ray data from two epochs, \pg14\ appeared to exhibit
two distinct states:
a variable, high-flux state (revolution 206); and a quiescent low-flux one 
(revolution 373).
At both epochs, a power law component with $\Gamma \approx 2.24$ was required,
while during the high-flux state an additional, steeper ($\Gamma \approx 2.5$) 
power law was also needed.
Significant spectral and flux variability was seen during the high-flux
state, whereas during revolution 373, the variability was negligible.

\item[(2)]
Short-term variability (within revolution 206) and long-term variability
(between the two epochs) was consistent with spectral hardening with
diminishing intensity.  In comparing the hard ($0.7-10\keV$) and
soft ($0.2-0.7\keV$) light curves during the high-flux state, there
appeared to be a short lag such that the hard energy band followed the soft by
about $3000\s$.

\item[(3)]
The radiative efficiency measurement during the bright, revolution 206 
observations was $\eta \gs 0.29$.  The value is consistent with emission from
an environment around a Kerr black hole or relativistic beaming.

\item[(4)]
Though not statistically significant, there appeared to be some
UV variability during the high state, which showed similar trends as the 
X-ray light curve.

\item[(5)]
The optical-to-X-ray spectral slope was estimated at both epochs from 
simultaneous UV data, and found to be significantly flatter during the
brighter X-ray phase.  
UV variability between the two epochs was within about $10\%$ and typical
for quasars, whereas the average broadband X-ray flux changed by about
$60\%$.

\end{itemize}

\subsection{The physical nature of the X-ray emission}
\label{sect:phnat}

The broken power law and blackbody plus power law models presented thus far 
are informative. 
For example, both models illustrated significant spectral variability between 
the two epochs, as well as on shorter time scales during revolution 206, but
they lack physical intuition.
A sudden break in a power law, or a high and variable blackbody temperature,
appears unphysical.  

In this section we attempt to derive a physically
motivated model for the spectral behaviour of \pg14.  In doing so, we demand
that all the different spectra can be fitted by a physically consistent
model.  This means that we not only attempt to describe the spectra at
both epochs with a consistent model, but also the high- and low-flux spectra
during revolution 206.  Consequently, the fits in this section include three
rather than two data sets (high-flux revolution 206; low-flux revolution 206;
lowest flux revolution 373).  For simplicity, we use only the pn data in this
section.

As opposed to a broken power law, a double power law does have physical
foundation. 
Initially we fit all three spectra with a double power law model.
We required that the photon indices were constant in each spectrum and allowed
only the normalisations to vary.  This provided a good fit to the data
($\chidof = 1.00/1607$) and illustrated how the hard power law component
dominated during the lowest flux state (revolution 373) and the steeper
component became more relevant with increasing flux.

We continued to build on this by also fixing the normalisation of the 
hard power law to be constant in all three spectra.  Therefore, the only
variable component at each epoch was the normalisation of the soft
power law.  This provided a slightly worse, but still acceptable
fit ($\chidof = 1.04/1609$).  The hard power law component 
($\Gamma = 2.24^{+0.02}_{-0.03}$) was essentially established by
the revolution 373 spectrum.  The contribution of the steep power law 
($\Gamma = 2.55\pm0.02$) 
component to the total spectrum during revolution 373 was negligible.
However, as the flux increased, the contribution of soft component became
more important.  In the highest flux state (high state during revolution 206)
the steep component was attributed to $62\pm1\%$ of the unabsorbed,
$0.25-10\keV$ flux, diminishing slightly to $58\pm1\%$ in the low-flux
spectrum of revolution 206, and completely negligible ($0\%$) at 
revolution 373.

It is interesting that this simple double power law model provides a reasonable
fit to three different spectra.  As seen throughout the paper, the soft
component was more variable on short and long time scales.
The fastest time scales at which we would expect to see variations
associated with the accretion disc, would be on dynamical time scales
($t_{d} = 2\times10^{3} M_8 (\frac{r}{2\rg})^{3/2} \s$).  For \pg14,
with a black hole mass of $\sim 10^9\Msun$ ($M_8=10$)
and $r=\rg=GM/c^2$ this corresponds
to variation on the order of $20\ks$ in the observed frame.  We are clearly
seeing variations on much shorter time scales.

Emission from an accretion disc could be consistent with the harder power
law component.  The absence of clear signatures from the disc
(e.g. iron emission line, reflection component above $10\keV$, and
variability) is not so surprising given the high luminosity
($>10^{46}\ergps$) of \pg14\ (e.g. see Reeves \& Turner 2000).
In fact, the harder power law in the double power law model can be 
successfully replaced with a thermal Comptonisation (Titarchuk \&
Mastichiadis 1994) component
($\chidof = 1.05/1607$).  In this case, the steep power law remained
unchanged compared to the double power law fit.  The seed photons
have a temperature of $kT\approx 33\eV$ and the plasma has a temperature 
of about $56\keV$.  The situation is not inconsistent with 
predicted physical conditions, although the measured model parameters are
not well constrained by the fit. 

It is conceivable that the soft, variable power law component can be attributed
to emission from an inner jet.  The slope of the X-ray spectrum 
($\Gamma \approx 2.5$) is steeper than the radio spectrum
(e.g Ulvestad \et 2005), which argues against inverse Compton
scattering as the emitting process.  The steep spectrum and remarkable 
variability strongly favour synchrotron radiation (e.g. Wilson \& Yang 2002).
 
The jet could be working intermittently:
present and dominant during the high-flux revolution 206 observation,
and absent during the later, low-flux observation.  In the spectral fits
presented in this section, we have demanded that the spectral slope
remains constant at all epochs.  This is, of course, not a necessary 
constraint as the X-ray slopes of jet emitting AGN are know to vary 
dramatically on various time scales.
In fact, based on spectral modelling, we do not rule out variations in
the photon index of the jet component between epochs, but note that the
lack of variability seen during the low-flux state, makes the argument 
pointless.
It should also be noted that the jet emission process is probably better
described by a curved spectrum in the \xmm\ energy range (Perlman \et 2005).  
Variations in this
curvature can be responsible for some of the spectral variation implied
by the difference spectrum and the flux-flux plot analysis
(Figure~\ref{fig:xdiffs} and \ref{fig:ff}, respectively).

Combined jet and accretion disc models have been fitted to a
number of nearby, radio-loud, unbeamed objects
(e.g. Cen~A, Evans \et 2004; NGC~4261, Zezas \et 2005), 
and have also been proposed to 
account for the apparently higher X-ray flux in beamed objects
(e.g. Browne \& Murphy 1987; Kembhavi 1993).  
The fact that neither component in \pg14\ appears to suffer from
obscuration, favours the conception that \pg14\ is beamed as well
(consistent with its radio jet).  If the presumable lag detected between
the two components (Figure~\ref{fig:ccf}) is attributed to light travel time
delays, then the two components (likely jet base and accretion disc corona)
are separated by only about $1\rg$.  This essentially places the base of the
jet inside the corona, giving credence to the idea that a jet could be
replenishing the plasma reservoir with high-energy particles.

\subsection{The spectral energy distribution }

The spectral and short-term timing behaviour do support the notion that 
\pg14\ is being viewed pole-on (Blundell \et 2003).
This can, in principle, describe the weakness of the high-ionisation
emission lines seen
in the optical and UV if, as Blundell \et describe, we are looking directly
into the broad line region.  On the other hand, the strong, low-ionisation
\feii\ emission could be collisionally produced at large distances where
the radio jet components are found ($\sim 100\pc$ from the core,
Blundell \et). 

The suggestion that the X-ray jet works intermittently also describes why
the SED is consistent with an unbeamed continuum, assuming that the
jet is more often off than on.  This follows from the dramatic differences
in $\alpha_{ox}$, as measured from simultaneous X-ray and UV data, which
indicate that the continuum can certainly be much flatter (apparently beamed)
when the jet is on.
In addition, if the X-rays are assumed to originate from the base of the jet
and hence feeding the radio jet, the irregularity of the X-ray jet emission
could help explain the weakness of the radio emission relative to the optical.

\section{Conclusions}

We have presented two \xmm\ observations, separated by eleven months,
of the radio-quiet quasar \pg14.  The X-ray behaviour is consistent as
arising from a combination of a relativistic jet and accretion disc processes.
The spectral and timing variability strongly suggest that the jet works
intermittently, which could help describe some of the complexities seen in
the SED of the quasar.  The variations found in 
$\alpha_{ox}$, measured from contemporaneous X-ray and UV data, demand the
need for simultaneous, multiwavelength observations to accurately define
the SED of \pg14.


\section*{Acknowledgements}

Based on observations obtained with \xmm, an ESA science mission with
instruments and contributions directly funded by ESA Member States and
the USA (NASA). 
LCG thanks Wolfgang Brinkmann, Iossif Papadakis and the referee for helpful
comments and discussion.



\bsp
\label{lastpage}
\end{document}